\documentclass{article}

\usepackage{arxiv}

\usepackage[utf8]{inputenc} 
\usepackage[T1]{fontenc}    
\usepackage{hyperref}       
\usepackage{url}            
\usepackage{booktabs}       
\usepackage{amsfonts}       
\usepackage{nicefrac}       
\usepackage{microtype}      
\usepackage{lipsum}		
\usepackage{graphicx}
\usepackage{natbib}
\usepackage{doi}

\RequirePackage{amsmath}
\RequirePackage{amssymb}
\RequirePackage{bbm}
\RequirePackage{bm}
\RequirePackage{xfrac}

\newcommand{\N}{\mathbb{N}}
\newcommand{\R}{\mathbb{R}}
\newcommand{\E}{\mathbb{E}}
\newcommand{\bbP}{\mathbb{P}}
\newcommand{\bbQ}{\mathbb{Q}}

\newcommand{\ind}{\mathbbm{1}}

\newcommand{\calX}{\mathcal{X}}

\newcommand{\calF}{\mathcal{F}}
\newcommand{\calB}{\mathcal{B}}
\newcommand{\calH}{\mathcal{H}}
\newcommand{\calI}{\mathcal{I}}
\newcommand{\calG}{\mathcal{G}}
\newcommand{\calM}{\mathcal{M}}

\newcommand{\calA}{\mathcal{A}}
\newcommand{\calP}{\mathcal{P}}

\newcommand{\frakF}{\mathfrak{F}}

\newcommand{\frakG}{\mathfrak{G}}
\newcommand{\frakH}{\mathfrak{H}}
\newcommand{\frakI}{\mathfrak{I}}

\newcommand{\bfH}{\mathbf{H}}

\newcommand{\bfS}{\mathbf{S}}

\newcommand{\bfdelta}{{\bm\delta}}
\newcommand{\bfpi}{{\bm\pi}}

\newcommand{\rmd}{\mathrm{d}}

\newcommand{\bdot}{~{\bm\cdot}~}

\newcommand{\myind}[1]{\ind_{\lbrace #1 \rbrace}}
\newcommand{\myE}[2]{\E_{#1}\left[ #2 \right]}
\newcommand{\myCond}[3]{\E_{#1}\left[\left. #2 \right\vert #3 \right]}
\newcommand{\mySet}[2]{\left\lbrace #1 : #2 \right\rbrace}
\newcommand{\ixA}{a}
\newcommand{\rndens}[2]{\frac{\mathrm{d} #1}{\mathrm{d} #2}}

\RequirePackage{amsthm}
\RequirePackage{thmtools}
\RequirePackage{thm-restate}
\declaretheorem[title=Theorem, style=plain]{thm}
\declaretheorem[numberlike=thm, title=Lemma, style=plain]{lem}
\declaretheorem[numberlike=thm, title=Corollary, style=plain]{cor}
\declaretheorem[title=Definition, style=definition]{defn}
\declaretheorem[title=Assumption, style=definition]{ass}
\declaretheorem[title=Remark, style=remark]{rem}
\declaretheorem[title=Example, style=remark]{exa}

\RequirePackage{enumitem}
\RequirePackage{hyperref}

\title{Large Platonic Markets with Delays}

\author{Yannick Limmer\thanks{Present Address:  Technische Universität München, Chair of Financial Mathematics, Parkring 11, 85748 Garching-Hochbrück, Germany.} \\
	Mathematics Institute\\
	Ludwig-Maximilians-Universität München\\
	Munich, Germany \\
	\texttt{y.limmer@tum.de} \\
	\And
	Thilo Meyer-Brandis \\
	Mathematics Institute\\
	Ludwig-Maximilians-Universität München\\
	Munich, Germany \\
	\texttt{meyer-brandis@math.lmu.de} \\
}

\renewcommand{\shorttitle}{Large Platonic Markets with Delays}

\hypersetup{
pdftitle={A template for the arxiv style},
pdfsubject={},
pdfauthor={Yannick Limmer, Thilo Meyer-Brandis},
pdfkeywords={arbitrage theory, large platonic financial markets, market delays},
}

\begin{document}
\maketitle

\begin{abstract}
    The objective is to develop a general stochastic approach to delays on financial markets. We suggest such a concept in the context of large platonic markets, which allow infinitely many assets and incorporate a restricted information setting. The discussion is divided into information delays and order execution delays. The former enables modeling of markets where the observed information is delayed, while the latter provides the opportunity to defer the indexed time of a received asset price. Both delays may be designed randomly and inhomogeneously over time. We show that delayed markets are equipped with a fundamental theorem of asset pricing and our main result is inheritance of the no asymptotic Lp-free lunch condition under both delay types. Eventually, we suggest an approach to verify absence of Lp-free lunch on markets with multiple brokers endowed with deviating trading speeds.
\end{abstract}

\keywords{arbitrage theory \and large platonic financial markets \and market delays}

\section{Introduction}\label{sec:Intro}
Modern financial markets are entirely digitized and therefore allow by means of technological advancements and state-of-the-art information processing techniques deviating trading paces amongst market participants. The difference in speed, however, is a direct violation of the assumption that all individuals have simultaneous access to information and quoted asset prices, as it is required for frictionless arbitrage theory. Hence, the necessity of developing a mathematical framework that incorporates explicit representations of delays in the information flow from a trader's viewpoint and a general approach to deferred order executions is an immediate consequence. In this article, we suggest such a concept.

Our approach is based on the considerations of \cite{CuchieroKleinTeichmann20}, where they extended C. Stricker's fundamental theorem of asset pricing (FTAP) \citep{Stricker90} to large financial markets in a restricted information setting. Thereby, \textit{large} indicates a possibly uncountable number of asset price processes, while \textit{restricted information} refers to the modelling assumption that the information available for trading is represented by a filtration smaller than the one portraying all market information. Two-filtration settings where first introduced to arbitrage theory by \cite{KabanovStricker06}, where they proved a FTAP for discrete-time markets under such conditions. Their result was then generalised to continuous-time markets by \cite{CuchieroKleinTeichmann16}, in the sense of F. Delbaen's and W. Schachermayer's FTAP \citep{DelbaenSchachermayer94}. The restricted information framework is here of particular interest due to the accessibility of modelling the information's availability in time.

We will employ the generality of this setting to establish two types of delays, the \textit{information delay} and the \textit{order execution delay}. The former characterizes the point in time when information for trading is available; the latter specifies the time lag between the placement of the order and its execution. We show that both types of delays can be incorporated in a general way into the framework. To be precise, we allow delays to be inhomogeneous and random, i.e. the dependence on the choice of asset, time, and event in the probability space is possible. As a result, a fully-fledged arbitrage theory becomes available for the resulting delayed markets.

This will lead to the main results of this article, namely the inheritance of absence of $L^p$-free lunch under market delays. The proof is quite straight forward for information delays; however, it requires additional effort for order execution delays. Such a result may be desirable in the following scenarios:
\begin{itemize}[label = --, wide]
    \item Market frictions arising from an unknown time of execution. For instance, suppose an agent issues a contingent claim but can purchase a replicating portfolio simultaneously for the same price. These actions constitute a perfect hedge, the overall position of the agent is zero with no exposure to the market. However, if the replicating portfolio's purchase is deferred or the information regarding its price is not up to date, a time lag may emerge between both transactions. During this period, the agent is exposed to market fluctuations: while already being invested in the contingent claim, the price of the replicating portfolio may vary over this time span.
    \item Market participants with deviating paces to process information. Such scenarios arise, for example, by the initially mentioned advancements in technology that are only accessible to some individuals. The framework established in this article provides an opportunity to determine the absence of $L^p$-free lunch in markets with such a complex time structure.
    \item Access through various brokers. Similar to the preceding, it is also possible to consider a scenario where the trader can choose between brokers to place her trade on the exchange. Here, a broker may be interpreted as an exchange's interface to execute trades. Brokers may differ in their pace of operation non-deterministically. 
    \item Speed bumps. Some exchange platforms add synthetic delay on the arrival or execution of trading orders to ensure that financial markets are a level playing field, as an answer to the unfairness resulting from high-frequency traders' dominance in speed. For a detailed discussion on the effectiveness of speed bumps in an economic context, the reader may, for instance, refer to \cite{BaldaufMollner20, BrolleyCimon20, Hoffmann14, KyleLee17, PagnottaPhilippon18}.
\end{itemize}

The remainder of the article is organized as follows. Section \ref{sec:Pre} briefly summarizes the results from \cite{CuchieroKleinTeichmann20}, Section \ref{sec:Info} covers information delays, and Section \ref{sec:Exec} is dedicated to order execution delays. In the latter section, we also derive a connection between both types of delays and briefly discuss the scenario of multiple brokers.

\section{Large platonic markets}\label{sec:Pre}
The objective of this section is to briefly summarise the underlying framework established in \cite{CuchieroKleinTeichmann20}. 

For this, set the initial time to $0$, fix any time horizon $T \in (0, \infty]$, and consider a filtered probability space $(\Omega, \calG, (\calG_{t})_{t \in [0,T]}, \bbP)$ as a stochastic basis. Here, $\Omega$ represents the state space, $\frakG := (\calG_{t})_{t \in [0,T]}$ the filtration, and $\bbP$ the physical measure which is defined on the $\sigma$-field $\calG := \calG_T$. Moreover, for every index $\ixA$ from an arbitrary -- countable or uncountable -- parameter set $I$, let $S^\ixA = (S^\ixA_t)_{t \in [0,T]}$ be a $\frakG$-adapted stochastic process.

Next, we introduce for every $A \subseteq I$ with $\vert A \vert < \infty$ a filtration $\frakF^A := (\calF^A_t)_{t \in [0,T]} \subseteq \frakG$, i.e. $\calF^A_t \subseteq \calG_t$ for all $0 \leq t \leq T$. In order to determine a reasonable nesting of these filtrations, we define an index system.
\begin{defn}[Index system]\label{def:IndexSystem}
A sequence of $I$-subset families, $(\calA^n )_{n \in \N}$ with $\calA = \bigcup_{n \in \N}\calA^n$, is called \textit{index system} if and only if
\begin{enumerate}[label = \roman*)]
    \item \label{cond:IndexSystem_A}for every $n \in \N$ it holds
    \begin{align*}
        \calA^n \subseteq \mySet{A \subseteq I}{\vert A \vert = n},
    \end{align*}
    \item \label{cond:IndexSystem_B}the \textit{refining property}  is fulfilled, i.e. $A_1, A_2 \in \calA$ yields $A_1 \cup A_2 \in \calA$, and 
    \item \label{cond:IndexSystem_C}for $A_1, A_2 \in \calA$ with $A_1 \subseteq A_2$ it follows that $\frakF^{A_1} \subseteq \frakF^{A_2}$. This is called \textit{monotonicity property} and is the case if and only if $\calF_t^{A_1} \subseteq \calF_t^{A_2}$ for all $t \in [0,T]$.
\end{enumerate}
\end{defn}
We now assume that $(\calA^n )_{n \in \N}$ with $\calA = \bigcup_{n \in \N}\calA^n$ is an index system that \textit{agrees} with the family of filtrations $(\frakF^A)_{A \subseteq I}$, meaning that property \ref{cond:IndexSystem_C} is satisfied. This simplifies the definition of trading, since it now can be done on small market fractions $A \in \calA$ and then generalized by the use of limits later on.

\begin{defn}[$\frakF^A$-simple trading]\label{def:SimpleTrading}
For $m \in \N$, denote by $\lbrace t_i \rbrace_{i = 0}^{m} \subseteq [0,T]$ an ordered, finite set of time points. Further, let $n\in \N$ and $A = \lbrace \ixA_1, \ldots, \ixA_n \rbrace \in \calA^n$.
\begin{enumerate}[label = \roman*)]
    \item\label{def:SimpleTrading_Strategy} An \textit{$\frakF^A$-simple trading strategy} $\bfH^A$ is a $\R^n$-valued process of the form 
    \begin{align*}
        \bfH^A = \sum_{i = 1}^{m} \bfH^A_{t_{i-1}} \ind_{(t_{i-1}, t_{i}]},
    \end{align*}
    where $\bfH^A_{t_{i-1}} = (H^{\ixA_1}_{t_{i-1}}, \ldots, H^{\ixA_n}_{t_{i-1}})^\intercal$ is a $\calF^A_{t_{i-1}}$-measurable random vector for every index $i \in \lbrace 1, \ldots, m \rbrace$. Furthermore, the strategy is called \textit{bounded} if $\sup_{i \in \lbrace 1, \ldots, m \rbrace, j \in \lbrace 1, \ldots, n \rbrace} \vert H^{\ixA_j}_{t_{i-1}} \vert < \infty$.
    \item\label{def:SimpleTrading_Wealth} The \textit{wealth process} obtained from an $\frakF^A$-simple bounded trading strategy is given by
    \begin{align*}
        \left(\bfS^A \bdot \bfH^A\right)_t = \sum_{j = 1}^{n}\sum_{i = 1}^{m} H^{\ixA_j}_{t_{i-1}}\ind_{(t_{i-1},1]} (t)
        \left( S^{\ixA_j}_{t_{i}\wedge t}- S^{\ixA_j}_{t_{i-1}}\right), \quad t \in [0,T],
    \end{align*}
    whereat $\bfS^A = (S^{\ixA_1}, \ldots, S^{\ixA_n})^\intercal$ is a sub-family of the above introduced asset price processes. 
    \item Therefore, the set
    \begin{align*}
        \calX^A = \mySet{(\bfS^A \bdot \bfH^A)_{t \in [0,T]}}{ \text{$\bfH^A$ $\frakF^A$-simple, bounded trading strategy}}
    \end{align*}
    contains all wealth processes in the small market determined by assets with indices from $A$ and $\calX^n = \bigcup_{A \in \calA^n} \calX^A$ consists of all wealth processes obtained by trading at most $n$ assets. These unite to $\calX = \bigcup_{n \in \N} \calX^n$ and the terminal values thereof are denoted by $K_0 := \lbrace X_T : X \in \calX\rbrace$.    
\end{enumerate}
\end{defn}
With that, the modelling of a financial market is concluded, and we will refer to a bundle $\left((S^\ixA)_{\ixA \in I}, \frakF, \frakG, T \right)$ -- where $(S^\ixA)_{\ixA \in I}$ is the family of asset price processes, $\frakF$ the collection of trading filtrations, $\frakG$ the general filtration, and $T$ denotes the maturity -- as \emph{market}. If these elements are defined as above, the market is called \textit{standard platonic}. 

\medskip
To conclude this section, we will repeat the fundamental theorem of asset pricing established in \cite{CuchieroKleinTeichmann20}. Henceforth, fix $p \in [1, \infty)$. In order to be able to work in an $L^p$-setting, we require the following assumption. 

\begin{ass}\label{ass:Lpintegrable}
The set of probability measures $\calP_p$, defined by
\begin{align*}
    \calP_p := \mySet{\bbP' \sim \bbP}{K_0 \subseteq L^p(\Omega, \calG, \bbP')},
\end{align*}
is assumed to be non-empty.
\end{ass}

With that, we are able to introduce a no-arbitrage criterion for this framework. For this, let $C := K_0 - L_+^0(\Omega, \calG, \bbP)$.

\begin{defn}[NAFLp]\label{def:NoAsymptoticLpFreeLunch}
A large platonic financial market fulfils the \emph{no asymptotic $L^p$-free lunch} condition if and only if for some $\bbP' \in \calP_p$ it holds
\begin{align}
    \overline{C_p(\bbP')} \cap L_+^p (\Omega, \calG, \bbP') = \lbrace 0 \rbrace \tag{NAFLp} \label{id:NoAsymptoticLpFreeLunch}
\end{align}
with $C_p(\bbP') := C \cap L^p(\Omega, \calG, \bbP')$. The closure is considered with respect to the $\Vert \cdot \Vert_{L^p(\Omega, \calG, \bbP')}$-norm topology. 
\end{defn}

In order to state the fundamental theorem of asset pricing, a suitable notion of equivalent martingale measures is needed. 

\begin{defn}[Notion of equivalent martingale measures]\label{def:EMM}
Let $q \in \lbrace 0 \rbrace \cup (1, \infty]$. The set $\calM^{q}$ contains all measures for which optional projections of asset price processes on finite index subsets $A \in \calA$ are $\frakF^{A}$-martingales and thereby have a $\bbP'$-density in $L^q(\Omega, \calG, \bbP')$ for some $\bbP' \in \calP_p$. To be precise,
\begin{enumerate}[label = \roman*)]
    \item for $q = 0$ define
    \begin{align*}
        \begin{split}
            \calM^0 := \big\lbrace \bbQ \sim \bbP ~:~ \forall \ixA \in A \text{ holds } \myCond{\bbQ}{S^\ixA_u}{\calF^A_t} = \myCond{\bbQ}{S^\ixA_t}{\calF^A_t} \text{ a.s.},&  \\ \text{ with } A \in \calA, 0 \leq t \leq u \leq T &\big \rbrace,
        \end{split}
    \end{align*}
    \item while for $\bbP' \in \calP_p$ and $q \in (1, \infty]$, where $q$ is the conjugate to $p$ and therefore fulfils $\sfrac{1}{p}+\sfrac{1}{q} = 1$, fix
    \begin{align*}
        \calM^q(\bbP') := \calM^0 \cap \Big\lbrace \bbQ \sim \bbP ~:~ \rndens{\bbQ}{\bbP'} \in L^q(\Omega, \calG, \bbP')\Big\rbrace. 
    \end{align*}
    and set $\calM^q := \bigcup_{\bbP' \in \calP_p}\calM^q(\bbP')$.
\end{enumerate}
\end{defn}

With that, we have all the necessary ingredients to repeat the main result of \cite{CuchieroKleinTeichmann20}.

\begin{thm}[Fundamental theorem of asset pricing, \cite{CuchieroKleinTeichmann20}]\label{thm:FTAP}
Assert Assumption \ref{ass:Lpintegrable} and let $p \in [1,\infty)$, $\bbP' \in \calP_p$. Then, the \eqref{id:NoAsymptoticLpFreeLunch} condition is satisfied (with respect to $\bbP'$) if and only if $\calM^q(\bbP') \neq \emptyset$, with $q \in (1 , \infty]$ being the conjugate of $p$.
\end{thm}

\section{Information delays}\label{sec:Info}
When a trade is ordered on a small market $S^A$, $A \in \calA$ at some time $t \in [0,T]$, the decision therefore is based on the information $\calF^A_t$. The main characteristic of an information delay is to point in this situation to a $\sigma$-field earlier in $\frakF^A$. To allow for inhomogeneous random shifts in time, we use the concept of stopping time processes.

\begin{defn}[Information delay]\label{def:InformationDelay}
The tuple $(\frakH, \delta)$, where $(\delta(t))_{t \in [0,T]}$ is a real valued random process on the filtered probability space $(\Omega, \calG, \frakG, \bbP)$ and $\frakH := (\calH_t)_{t \in [0,T]}$ a filtration with $\frakH \subseteq \frakG$, is called \emph{information delay} (or \emph{$\delta$-delay}) if 
\begin{enumerate}[label = \roman*)]
	\item for any $t \in [0,T]$, $\delta(t)$ is a stopping time with respect to $\frakH$, that is $\lbrace \delta(t) \leq s \rbrace \in \calH_s$ for all $s \in [0,T]$;
	\item the boundaries $0 \leq \delta(t) \leq t$ are met for all $t \in [0,T]$; and the process is 
	\item path-wise monotone, i.e. for all $0 \leq t \leq s \leq T$ and $\omega \in \Omega$ holds $\delta(t, \omega) \leq \delta(s, \omega)$.
\end{enumerate}
\end{defn}

In the above definition, boundaries, as well as path-wise monotonicity, ensure well-definedness of the resulting \textit{delayed filtration}. For a small market, this is defined and expressed in the following lemma. 

\begin{lem}[$\delta$-delayed trading filtration]\label{lem:DeltaDelayedTradingFiltration}
	Let $(\frakH, \delta)$ be an information delay. If $\frakH \subseteq \frakF^A$ for a fixed $A \in \calA$ is satisfied, then the \emph{$\delta$-delayed trading filtration} $\frakF^{A, \delta} := (\calF^{A, \delta}_t)_{t \in [0,T]}$ specified by 
	\begin{align*}
	\calF^{A, \delta}_t := \calF^A_{\delta(t)},\quad t \in [0,T]
	\end{align*}
	is well-defined. Here, $\calF^A_{\delta(t)}$ denotes the $\sigma$-field of the $\delta(t)$-past.
\end{lem}
\begin{proof}
Note that $\frakH \subseteq \frakF^A$ guarantees that for $t \in [0,T]$, $\delta(t)$ is a $\frakF^A$-stopping time. Since the $\sigma$-field of the $\delta(t)$-past is indeed a $\sigma$-field, the monotonicity property of $\delta$ readily ensures that $\frakF^{A, \delta}$ is in fact a filtration. 
\end{proof}

In order to obtain full generality, we intend to equip every market portion with an individual information delay. We have to guarantee that this agrees with the construction of large financial markets via the index system in Definition \ref{def:IndexSystem}.

\begin{defn}[Large $\delta$-delayed markets]\label{def:LargeDeltaDelayedMarkets}
Let $\bfdelta := (\delta^A)_{A \in \calA}$ be a family of information delays, where for every $A \in \calA$ the delay information $\frakH^A$ corresponding to $\delta^A$ satisfies $\frakH^A \subseteq \frakF^A$. Then, the family of filtrations $(\frakF^{A, \bfdelta})_{A \in \calA}$ is defined recursively for $A \in \calA$ via
\begin{align*}
    \frakF^{A,\bfdelta} := (\calF^{A,\bfdelta}_t)_{t \in [0,T]} \quad \text{with} \quad 
    \calF^{A,\bfdelta}_t := \sigma \Big( \calF^{A, \delta^A}_t \cup \bigcup_{A \supsetneq A' \in \calA} \calF^{A',\bfdelta}_t \Big), \quad \forall t \in [0,T].
\end{align*}
\end{defn}

\begin{lem}\label{lem:LargeDeltaDelayedMarkets}
The family of filtrations $(\frakF^{A, \bfdelta})_{A \in \calA}$ agrees with the index system $\calA$.
\end{lem}
\begin{proof}
    The structure of the proof is the following: initially \ref{step:LDDM_A} we show that $\frakF^{A,\bfdelta}$ is in fact a filtration and thereafter \ref{step:LDDM_B} demonstrate that the monotonicity property of the index set is fulfilled. 
    \begin{enumerate}[label = \roman*), wide]
       \item \label{step:LDDM_A} We argue by induction over $\vert A \vert$ for $A \in \calA$. The case $\vert A \vert = 1$ is a direct consequence of Lemma \ref{lem:DeltaDelayedTradingFiltration}. 
       
       For the induction step, let $ 0  \leq t \leq u \leq T$ and
       \begin{align*}
           F \in \calF^{A, \delta^A}_t \cup \bigcup_{A \supsetneq A' \in \calA} \calF^{A',\bfdelta}_t \quad.
       \end{align*}
       Lemma \ref{lem:DeltaDelayedTradingFiltration} already ensures that $\calF^{A, \delta^A}_t \subseteq \calF^{A, \delta^A}_u$ and by induction hypothesis we have 
       \begin{align*}
           \bigcup_{A \supsetneq A' \in \calA} \calF^{A',\bfdelta}_t \subseteq \bigcup_{A \supsetneq A' \in \calA} \calF^{A',\bfdelta}_u.
       \end{align*}
       Therefore, the generating system of $\calF^{A, \bfdelta}_t$ is a subset of the one generating $\calF^{A, \bfdelta}_u$, hence $\calF^{A, \bfdelta}_t \subseteq \calF^{A, \bfdelta}_u$.
       \item \label{step:LDDM_B}Fix any $t \in [0,T]$ and $A, A' \in \calA$ with $A \subsetneq A'$. We have to demonstrate that $\calF^{A,\bfdelta}_t \subseteq \calF^{A',\bfdelta}_t$, what is trivially fulfilled since $\calF^{A, \bfdelta}_t$ is contained within the generating set of $\calF^{A', \bfdelta}_t$.
    \end{enumerate}
   
    We have shown the desired result.
\end{proof}

\begin{rem}
The above lemma allows us to consider any large market where market fractions are equipped with suitable information delays. However, due to the generality of the result, it is not straightforward to obtain a modelling procedure that guarantees a stalled flow of information on a market portion since delays on super index sets may overwrite the available information on smaller fractions with shorter delays. Therefore, it is reasonable to model the family of information delays $\bfdelta := (\delta^A)_{A \in \calA}$ such that 
\begin{align*}
    \delta^{A}(t) \geq \delta^{A'}(t), \quad \forall t \in [0,T], \quad \forall A,A' \in \calA \text{ with } A \subseteq A'. 
\end{align*}
In prose, this means that the delay increases with the size of the set. Consequently, modelling information is getting more accessible: While the overall flow of information on a bigger market portion is slow, some subsets may exist where faster information is available. The modelling of deviating trading paces is demonstrated in the following example.
\end{rem}

\begin{exa}[Modelling information delay]\label{exa:ModelInfoDelay}
Consider $I = \lbrace 1, 2, 3, 4 \rbrace$. We want to allow trading on the first two assets with fast information $\mathrm{FAST} := \lbrace 1, 2 \rbrace$ and trading on the third with slow information $\mathrm{SLOW} := \lbrace 3 \rbrace$. We combine both with $\mathrm{COMB} := \lbrace 1, 2, 3\rbrace$, to be conform with the index system. Eventually, we say that trading is done super slow on the entire index set, $\mathrm{SSLW} := I$. Choosing any suitable (non-delayed) filtrations, we note that $\lbrace \mathrm{FAST}, \mathrm{SLOW}, \mathrm{COMB}, \mathrm{SSLW} \rbrace$ is in fact an index system. We now introduce a family of information delays $\bfdelta := \lbrace \delta^{\mathrm{FAST}}, \delta^{\mathrm{SLOW}}, \delta^{\mathrm{COMB}}, \delta^{\mathrm{SSLW}} \rbrace$ with 
\begin{align*}
    \delta^{\mathrm{FAST}} > \delta^{\mathrm{SLOW}} = \delta^{\mathrm{COMB}} > \delta^{\mathrm{SSLW}}.
\end{align*}
The result is that on $\frakF^{\mathrm{SSLW}, \bfdelta}$ the information flow of assets 1 and 2 is fast, slower reported for the asset 3, and super slow information is received for asset 4.
\end{exa}
\begin{rem}
    It is not possible to "hide" information from trading on other assets, meaning that for $A, A' \in \calA$ and a family of information delays $\bfdelta$, the information $\frakF^{A,\bfdelta}$ is available for trading on $A'$ and vice versa. 
\end{rem}
\medskip

Our objective is now to show that the \eqref{id:NoAsymptoticLpFreeLunch} is preserved when adding an information delay to a standard platonic market. For this, we need the following auxiliary result, which states that the delayed filtration is coarser than the original.

\begin{lem}\label{lem:DelayReducesInformation}
For any $A \in \calA$, $t \in [0,T]$, and a family of information delays $\bfdelta = (\delta^A)_{A \in \calA}$, it holds that $\calF^{A, \bfdelta}_t \subseteq  \calF^{A}_t$.
\end{lem}
\begin{proof}
The proof is done by induction. It is clear that $\calF^{A, \bfdelta}_t \subseteq  \calF^{A}_t$ for $A \in \calA^1$, since $\calF_t^{A, \delta^A} \subseteq \calF^A_t$ due to the order of stopped filtrations and $\mySet{A' \in \calA}{A' \subsetneq A} = \emptyset$.

Now assume that the statement holds for all $A' \in \bigcup_{k = 1}^{n-1} \calA^k$ with $n \in \N$. This yields that 
\begin{align*}
    \calF^{A, \delta^A}_t \cup \bigcup_{A \supsetneq A' \in \calA} \calF^{A',\bfdelta}_t  \subseteq \calF^{A}_t \cup \bigcup_{A \supsetneq A' \in \calA} \calF^{A'}_t = \calF^A_t,
\end{align*}
since $\calF^{A, \delta^A}_t \subseteq \calF^A_t$ and $\calF^{A'}_t \subseteq \calF^{A}_t$ for all $A \supsetneq A' \in \calA$. The $\sigma$-field generated by a family of sets is the smallest $\sigma$-field containing all of these, hence $\calF^{A, \bfdelta}_t \subseteq \calF^A_t$.
\end{proof}

\begin{thm}[$\mathrm{NAFLp}$ with information delay]\label{thm:NAFLpWithInfoDelay}
	Consider a standard platonic market $((S^\ixA)_{\ixA \in I}, \frakF, \frakG, T)$ that satisfies the \eqref{id:NoAsymptoticLpFreeLunch} condition. Then, in the market $((S^\ixA)_{\ixA \in I}, \frakF^\bfdelta, \frakG, T)$, where the family of delayed filtrations $\frakF^\bfdelta$ corresponds to some family of information delays $\bfdelta$, asymptotic $L^p$-free lunch is absent. 
\end{thm}
\begin{proof}
    The standard platonic market $((S^\ixA)_{\ixA \in I}, \frakF, \frakG, T)$ satisfies the \eqref{id:NoAsymptoticLpFreeLunch} condition, hence we can invoke Theorem \ref{thm:FTAP} to conclude that there is a measure $\bbQ$ with $\rmd\bbQ/\rmd\bbP \in L^q(\Omega, \calG, \bbP)$ and
    \begin{align*}
        \forall \ixA \in A \text{ holds } \myCond{\bbQ}{S^\ixA_{u}}{\calF^A_t} = \myCond{\bbQ}{S^\ixA_{t}}{\calF^A_t} \text{ a.s.} \text{ with } A \in \calA, 0 \leq t \leq u \leq T.
    \end{align*}
    Fix any $0 \leq t \leq u \leq T$ and $A \in \calA$. We compute for $\ixA \in A$
    \begin{align*}
        \myCond{\bbQ}{S^\ixA_{u}}{\calF^{A, \bfdelta}_t}& =
        \myCond{\bbQ}{\myCond{\bbQ}{S^\ixA_{u}}{\calF^{A}_t}}{\calF^{A, \bfdelta}_t} \\
        & = \myCond{\bbQ}{\myCond{\bbQ}{S^\ixA_{t}}{\calF^{A}_t}}{\calF^{A, \bfdelta}_t} 
        = \myCond{\bbQ}{S^\ixA_{t}}{\calF^{A, \bfdelta}_t},
    \end{align*}
    where we asserted that $\calF^{A, \bfdelta}_t \subseteq  \calF^{A}_t$. This holds due to Lemma \ref{lem:DelayReducesInformation}, and we obtain a martingale measure $\bbQ$ for the market $((S^\ixA)_{\ixA \in I}, \frakF^\bfdelta, \frakG, T)$ in the sense of Definition \ref{def:EMM}. Another application of the fundamental theorem of asset pricing (Theorem \ref{thm:FTAP}), which is available due to Lemma \ref{lem:LargeDeltaDelayedMarkets}, yields the desired result.
\end{proof}

Theorem \ref{thm:NAFLpWithInfoDelay} coincides with our understanding of information flow on financial markets. If only an additional delay on the information's reporting speed is imposed, the thereby resulting market remains without the existence of an asymptotic $L^p$-free lunch. To put this another way, an asymptotic $L^p$-free lunch strategy in the market  $((S^\ixA)_{\ixA \in I}, \frakF^\bfdelta, \frakG, T)$ characterizes as well such strategy in the original market $((S^\ixA)_{\ixA \in I}, \frakF, \frakG, T)$. 

The converse of Theorem \ref{thm:NAFLpWithInfoDelay} does not hold in general, meaning that a market with information delay $((S^\ixA)_{\ixA \in I}, \frakF^\bfdelta, \frakG, T)$ does not necessarily remain free of asymptotic $L^p$-free lunch strategies if the information delay is omitted. 

\begin{exa}[Insider trading]\label{exa:InsiderTrading}
    We will briefly construct a well-known default example for insider trading to contradict the converse of Theorem \ref{thm:NAFLpWithInfoDelay}. For this, fix some constant $0 < h < 1$ and consider a market with one asset, $S := (S_t)_{t \in [0,1]}$, where $S_t := B_t$ for all $t \in [0,1]$ and $B := (B_t)_{t \in [0,1+h]}$ is a $\bbP$-Brownian motion. The maturity is set to $T = 1$. As a next step, we model the flow of information. The filtration $\frakG$ is determined by $\calG_t := \sigma(\lbrace B_{u + h} \rbrace_{u \in [0,t]})$ for all $t \in [0,1]$, while the trading information $\frakF$ is defined via $\calF_t := \sigma(\lbrace B_{u + h}\rbrace_{u \in [0,t]})$ for $t \in (0,1]$ and $\calF_0 = \sigma(B_0)$.  
    
    Furthermore, we introduce the information delay $\delta := (\delta(t))_{t \in [0,1]}$ with $\delta(t) := (t - h)_+$ for all $t \in [0,1]$. Accordingly, the $\delta$-delayed $\sigma$-field $\frakF^\delta$ is characterized by $\calF_t^\delta := \calF_{(t - h)_+} = \sigma(\lbrace B_{u}\rbrace_{u \in [0,t]})$ for all $t \in [0,1]$ due to Lemma \ref{lem:DeltaDelayedTradingFiltration}. It is obvious that on the market $(S, \frakF^\delta, \frakG, 1)$ we have $\bbP \in \calM^\infty(\bbP)$, hence Theorem \ref{thm:FTAP} yields that the $(\mathrm{NAFL}_1)$ condition is fulfilled. 
    
    However, there is a free lunch in the market $(S, \frakF, \frakG, 1)$. For $0 < t < 1 - h $, we have that
    \begin{align*}
        \calF_t \supseteq \sigma(\lbrace S_u \rbrace_{u \in [0,t]}) \cup \sigma(S_{t+h}),
    \end{align*}
    hence the terminal value of the wealth process constructed by $m = 1$, $t_0 = t$, $t_1 = t+h$, and $H_{t_1} := 2\myind{S_{t + h} \geq S_t} - 1$ is equal to $\vert S_{t+h} - S_t \vert$ and constitutes an $L^p$-free lunch.
\end{exa}

\section{Order execution delays}\label{sec:Exec}
Up to this point, we assumed that every trade ordered at time $t \in [0, T]$ is as well executed at this point in time. The objective is now to relax this assumption by introducing a random time lag between those actions. Since we intend to distinguish the delays for differing times and events, this will be done again by the use of stopping time processes. 

Since the delayed time may be after maturity, we henceforth consider assets $\lbrace S^\ixA \rbrace_{\ixA \in I}$ and the filtration $\frakG$ to be defined on the entire positive half-line $\R_+ \cup \lbrace \infty \rbrace$. Moreover, we need an additional assumption to the framework from Section \ref{sec:Pre} to ensure the well-definedness of stopped processes. 

\begin{ass} \label{ass:Progressive}
For any $\ixA \in I$, the price process $S^\ixA$ is asserted to be progressively measurable. Moreover, $S^\ixA$ is assumed to have a last element $S^\ixA_\infty \in \calG_\infty$. 
\end{ass}
\begin{defn}[Order execution delay]\label{def:ExecutionDelay}
	Let $\pi := (\pi(t))_{t \in [0,T]}$ be a real-valued random process on the filtered probability space $(\Omega, \calG, \frakG, \bbP)$ and $\frakI := (\calI_t)_{t \in \R_+}$ be a filtration satisfying $\frakI \subseteq \frakG$. Then, a tuple $(\frakI, \pi)$ is called \emph{order execution delay} (in brief \emph{execution delay} or \emph{$\pi$-delay}) if 
\begin{enumerate}[label = \roman*)]
	\item for any $t \in [0,T]$, $\pi(t)$ is a stopping time, that is $\lbrace \pi(t) \leq u \rbrace \in \calI_u$ for all $u \in \R_+$;
	\item the boundary $t \leq \pi(t)$ is met for all $t \in [0,T]$; and the process is 
	\item \label{cond:Pi-Monotone} monotone, i.e. for all $0 \leq t \leq u \leq T$ it holds $\pi(t) \leq \pi(u)$.
\end{enumerate}
An order execution delay is referred to as 
\begin{enumerate}[label = \roman*), resume]
    \item \emph{(right-)continuous} if $\pi(\cdot)$ is path-wise (right-)continuous and 
    \item \emph{strictly monotone} if in \ref{cond:Pi-Monotone} $\pi(t) < \pi(u)$ for all $0 \leq t < u \leq T$.
\end{enumerate}
For simplicity of notation, $(\frakI, \pi)$ is identified with $\pi$, whenever the relation to the \emph{delay information} $\frakI$ is clear or $\frakI$ is chosen arbitrarily. A family of execution delays is denoted by $\bfpi := (\pi^\ixA)_{\ixA \in I}$ and the $A$-indexed vector is referred to as $\pi^A := (\pi^{\ixA_1}, ..., \pi^{\ixA_n})$ for $A = \lbrace \ixA_1, \ldots, \ixA_n \rbrace \in \calA, n \in \N$.
\end{defn}

The concept of order execution delays is now used to redefine the market's asset price processes. 

\begin{defn}[Delayed asset price process and market]\label{def:DelayedAssetPriceProcess}
For a family of execution delays $\bfpi$, the \emph{delayed asset price processes} are defined by
\begin{align}
	\bfS^{A, \pi^A}_t := \left(\bfS^{\ixA_1}_{\pi^{\ixA_1}(t)}, \ldots, \bfS^{\ixA_n}_{\pi^{\ixA_n}(t)} \right) \quad \forall t \in [0,T], \quad \forall A = \lbrace \ixA_1, \ldots, \ixA_n \rbrace \in \calA. \label{id:DelayedAsset}
\end{align}
Moreover, the resulting market is denoted by $((S^{\ixA, \pi^\ixA})_{\ixA \in I}, \frakF, \frakG^\bfpi, T)$, where 
\begin{align*}
    \frakG^\bfpi := (\calG^\bfpi_t)_{t \in [0,T]}; \quad \calG^\bfpi_t  := \sigma\Big( \bigcup_{\ixA \in I} \calG_{\pi^\ixA(t)}\Big), \quad t \in [0,T].
\end{align*}
\end{defn}
\begin{rem}
    \begin{enumerate}[label = \roman*), wide]
        \item Since the index of $\pi$ is redundant, we simply write $\bfS^{A, \pi^A} \equiv \bfS^{A, \bfpi}$.
        \item For $\ixA \in I$, it is straightforward to see that $\pi^\ixA$ is $\calG^\bfpi$-adapted. Moreover, we say the execution delay $\pi^\ixA$ is \emph{progressive} if the stochastic process $\pi^\ixA$ is progressively measurable with respect to the filtration $\frakG^\bfpi$.
    \end{enumerate}
\end{rem}
\medskip 

As a next step, we investigate how execution delayed markets can be classified in previously discussed markets. For this, we begin with the following canonical observation that allows us to access all results from Section \ref{sec:Pre}, inter alia the fundamental theorem of asset pricing.

\begin{lem}[$\pi$-delayed trading]\label{lem:PiDelayedTrading}
    Let Assumption \ref{ass:Progressive} be fulfilled. Then, the market $((S^{\ixA, \bfpi^\ixA})_{\ixA \in I}, \frakF, \frakG^\bfpi, T)$ is a standard platonic market in the framework established in Section \ref{sec:Pre}.
\end{lem}
\begin{proof}
Assumption \ref{ass:Progressive} yields that the stochastic process $S^{\ixA, \bfpi}$ is well defined since an $\frakI$-stopping time is as well a $\frakG$-stopping time. The filtration $\frakG^\bfpi$ is obviously well defined and $S^{\ixA, \bfpi}$ is clearly $\frakG^\bfpi$ measurable. The remainder is trivial since the framework established in Section \ref{sec:Pre} does not impose any restrictions on the asset price processes. 
\end{proof}

As a next step, we will relate the notion of order execution delays to the previously established concept of information delays.

\begin{thm}\label{thm:RelationInformationExecutionDelay}
Assume that $\lbrace \ixA \rbrace \in \calA$ for all $\ixA \in I$ and assert Assumption \ref{ass:Progressive}. Let $\bfpi = (\pi^\ixA)_{\ixA \in I}$ be a family of continuous execution delays where $\frakI^\ixA$, the delay information with respect to $\pi^\ixA$, is contained in $\frakF^A$ whenever $\ixA \in A$ for all $A \in \calA$. Moreover, assume that for any $A \in \calA$ the trading filtration $\frakF^A$ is right-continuous. 

Then, the information in the market $((S^{\ixA, \bfpi})_{a \in I}, \frakF, \frakG^\bfpi, T)$ can be represented via an information delay on the market $((S^{\ixA, \bfpi})_{a \in I}, \Tilde{\frakF}, \frakG^\bfpi, T)$, where $\tilde{\frakF} := (\tilde{\frakF}^A)_{A \in \calA}$ are the (non-delayed) trading filtrations described by   
\begin{align*}
    \tilde{\frakF}^A := ( \Tilde{\calF}^A_{t} )_{t \in [0,T]} := \bigg( \sigma \Big( \bigcup_{\ixA \in A} \calF^A_{\pi^\ixA(t)} \Big) \bigg)_{t \in [0,T]}, \quad \forall A \in \calA.
\end{align*} 
\end{thm}

\begin{proof}
The family of filtrations $\Tilde{\frakF}$ is well-defined, since $\pi^\ixA(t)$ is a $\frakF^A$-stopping time for $\ixA \in A$ and all $A \in \calA$, $t \in [0,T]$. Moreover, it is straightforward to see that the market $((S^{\ixA, \bfpi})_{a \in I}, \Tilde{\frakF}, \frakG^\bfpi, T)$ is a standard platonic market due to Lemma \ref{lem:PiDelayedTrading} and the fact that for all $A \in \calA$, $t \in [0,T]$ we have $\Tilde{\calF}^A_t \subseteq \calG^\bfpi_t$. Note that Lemma \ref{lem:PiDelayedTrading} is available due to Assumption \ref{ass:Progressive}.

The assertion $\lbrace \ixA \rbrace \in \calA$ for all $\ixA \in I$ allows us to perform our analysis on a fixed $\ixA \in \calA$ and afterwards extend our result to bigger index-subsets. Thus, fix an arbitrary $\ixA \in I$ and for the sake of notation, write $\pi \equiv \pi^\ixA$.

We now introduce a candidate $\delta^a$ (henceforth $\delta \equiv \delta^a$) for the information delay by defining for fixed $\omega \in \Omega$
\begin{align*}
    \delta(t,\omega) := \pi^{-1}(t, \omega), \quad \forall t \in \R_+ \cup \lbrace \infty \rbrace,
\end{align*}
whereby we identify $\pi^{-1}(t, \omega) := \inf\mySet{s \in [0,T]}{\pi(s, \omega) \geq t} \wedge T$. Due to continuity, it is possible to interchange the infimum with a minimum in the preceding expression. If $\delta$ is an information delay, we can compute
\begin{align*}
    \tilde{\frakF}^{\lbrace \ixA \rbrace, \bfdelta} = \tilde{\frakF}^{\lbrace \ixA \rbrace, \delta} = ( \Tilde{\calF}^{\lbrace \ixA \rbrace}_{\delta(t)} )_{t \in [0,T]} =({\calF}^{\lbrace \ixA \rbrace}_{\pi(\pi^{-1}(t))})_{t \in [0,T]} = ({\calF}^{\lbrace \ixA \rbrace}_{t})_{t \in [0,T]},
\end{align*}
where we used the fact that $\pi \circ \pi^{-1}$ constitutes the identity mapping due to monotonicity and continuity.

For the above being true, we will show that $\delta$ is in fact an information delay.\footnote{Since we did not impose any boundaries on the elements of $\pi$, it is necessary for $\delta$ to be defined on the entire half-line $\R_+ \cup \lbrace \infty \rbrace$. As for markets, this natural extension of information delays follows without further ado.} Regarding Definition \ref{def:InformationDelay}, we need to prove that for fixed $t \in \R_+ \cup \lbrace \infty \rbrace$, $\delta(t)$ is an $\Tilde{\frakF}^{\lbrace \ixA \rbrace}$-stopping time, since boundedness and path-wise monotonicity are already ensured by definition. Consider any $s \in [0,T]$, so the objective is to demonstrate 
\begin{align*}
    \lbrace \delta(t) \leq s \rbrace \in \Tilde{\calF}^{\lbrace a \rbrace}_{s}. 
\end{align*}
First, we notice that
\begin{align}
    \Tilde{\calF}^{\lbrace a \rbrace}_{s} = \calF^{\lbrace \ixA \rbrace}_{\pi(s)} :=\mySet{F \in \calF_T^{\lbrace \ixA \rbrace}}{ \{\pi(s) \leq u\} \cap F \in \calF_u^{\lbrace \ixA \rbrace}{\text{ for all }} u \in [0,T]}. \label{id:calGPiS}
\end{align}
Next, we see that
\begin{align*}
    \lbrace \delta(t) \leq s \rbrace  = \lbrace \pi \circ \pi^{-1}(t) \leq \pi(s) \rbrace= \lbrace t \leq \pi(s) \rbrace = \lbrace \pi(s) < t \rbrace^c. 
\end{align*}
By the assumption that $\frakI^\ixA \subseteq \frakF^{\lbrace \ixA \rbrace}$, we know that $\lbrace \pi(s) \leq t \rbrace \in \calF^{\lbrace \ixA \rbrace}_t$. Accordingly, $\frakF^{\lbrace \ixA \rbrace}$ being right-continuous indicates $\lbrace \pi(s) < t \rbrace \in \calF^{\lbrace \ixA \rbrace}_t$ and therefore $\lbrace \pi(s) < t \rbrace \in \calF^{\lbrace \ixA \rbrace}_T$. Now, for arbitrary $0 \leq u < t$,
\begin{align*}
    \lbrace \pi(s) < t \rbrace \cap \lbrace \pi(s) \leq  u \rbrace = \lbrace \pi(s) \leq u \rbrace \in \calF^{\lbrace \ixA \rbrace}_u
\end{align*}
and for $t \leq u \leq T$, the above yields
\begin{align*}
    \lbrace \pi(s) < t \rbrace \cap \lbrace \pi(s) \leq  u \rbrace = \lbrace \pi(s) < t \rbrace \in \calF^{\lbrace \ixA \rbrace}_t \subseteq \calF^{\lbrace \ixA \rbrace}_u.
\end{align*}
Thus, $\lbrace \delta(t) \leq s \rbrace \in \Tilde{\calF}^{\lbrace \ixA \rbrace}_s$ due to \eqref{id:calGPiS}. 

It is left to treat the preceding for bigger index-subsets in $\calA$. For an arbitrary $A \in \calA$, fix the information delay $\delta^A := \min_{\ixA \in A} \delta^a$. We see that $\delta^A$ defines a proper information delay, also due to the fact that the maximum of finitely many stopping times characterizes itself a stopping time and that every $\frakF^{\lbrace \ixA \rbrace}$-stopping time is as well an $\frakF^A$-stopping time for $\ixA \in A$. We proceed by induction and assume that $\tilde{\frakF}^{A, \bfdelta} = \frakF^A$ for $A \in \bigcup_{k = 1}^{n-1} \calA^k$ with $n \in \N$. 

Fix now an $A \in \calA^n$ and compute for some fixed $t \in [0,T]$,
\begin{align*}
    \Tilde{\calF}^{A, \bfdelta}_t = \sigma \Big( \Tilde{\calF}^{A, \delta^A}_t \cup \bigcup_{A \supsetneq A' \in \calA} \Tilde{\calF}^{A',\bfdelta}_t \Big) 
    = \sigma \Big( \sigma \big( \bigcup_{\ixA \in A} \calF^A_{\pi^\ixA(\delta^A(t))} \big)\cup \bigcup_{A \supsetneq A' \in \calA} \calF^{A'}_t \Big).
\end{align*}
We are done as soon as we have established that $\sigma(\bigcup_{\ixA \in A} \calF^A_{\pi^\ixA(\delta^A(t))}) = \calF^A_t$. Observe that 
\begin{align*}
    \pi^\ixA(\delta^A(t)) = \pi^\ixA(\textstyle\min_{\ixA' \in A}\delta^{\ixA'}(t)) = \pi^\ixA(\textstyle\min_{\ixA' \in A}(\pi^{\ixA'})^{-1}(t)) \leq t,
\end{align*}
so $\bigcup_{\ixA \in A} \calF^A_{\pi^\ixA(\delta^A(t))} \subseteq \calF^A_t$ holds due to the order of stopped filtrations. For the other inclusion, assume $F \in \calF^A_t$ and define 
\begin{align*}
    \Pi^\ixA := \mySet{\omega \in \Omega}{\delta^\ixA(t) = \textstyle\min_{\ixA' \in A}\textstyle\delta^{\ixA'}(t)} \subseteq \Omega, \quad \forall \ixA \in A.
\end{align*}
Fix any $\ixA \in A$. It is a straightforward observation that $\Pi^\ixA \in \calF^A_t$ due to the boundedness of $\delta^\ixA(t)$ from above by $t$.  This directly implies that $F \cap \Pi^\ixA$ is $\calF^A_t$-measurable accordingly. However, we know by the definition of the stopped $\sigma$-field that
\begin{align*}
    (F \cap \Pi^\ixA) \cap \lbrace \pi^\ixA(\delta^A(t)) = t \rbrace \in \calF^A_{\pi^\ixA(\delta^A(t))},
\end{align*}
which reduces to $(F \cap \Pi^\ixA) \in \calF^A_{\pi^\ixA(\delta^A(t))}$. Since we can rewrite $F = \bigcup_{\ixA \in A} F \cap \Pi^\ixA$, we have shown $\tilde{\frakF}^{A, \bfdelta} = \frakF^A$.

We conclude that the information on the market $((S^{\ixA, \bfpi})_{a \in I}, \frakF, \frakG^\bfpi, T)$ may indeed be given via the family of information delays $\bfdelta = (\delta^A)_{A \in \calA}$ on the market $((S^{\ixA, \bfpi})_{a \in I}, \Tilde{\frakF}, \frakG^\bfpi, T)$.
\end{proof}
\medskip

The goal of the remainder is to provide an answer to whether a market with absence of free lunch remains free thereof if an execution delay is imposed. Primarily, we are interested in the conditions that are necessary for this result.

The first aspect is straightforward. Order execution delays allow per definition trades after maturity in the original market since the asset price indexed at time $T$ may be delayed to some $\pi(T) > T$. Hence, it is canonical to require that free lunch is absent in the original market at times after maturity.

The second facet is a technicality that is inevitable for applying Doob's optional sampling theorem, which is used to deduce the (conditional) expectation of stopped processes and allows us to prove martingale properties of execution delayed asset price processes under a suitable measure. The assertion reads as follows.

\begin{ass}\label{ass:Cadlag}
    For every $\ixA \in I$, $S^\ixA$ is a càdlàg stochastic process. 
\end{ass}

Eventually, we have to guarantee that the delay information is available for the agent. To put this another way, the agent may not be aware of the delay's magnitude at the time of trading; nonetheless, the delay is observable for her, i.e. she knows the time of execution once the order is executed. 

With that, we can state and prove the main result of this passage.

\begin{thm}[NAFLp with order execution delay]\label{thm:NAFLpWithExecDelay}
    Assert that Assumptions \ref{ass:Progressive} and \ref{ass:Cadlag} hold. Let $\bfpi = (\pi^\ixA)_{\ixA\in I}$ be a family of execution delays and the filtration $\frakI^\ixA$ corresponding to $\pi^\ixA$ satisfies $\frakI^\ixA \subseteq \frakF^A$ whenever $\ixA \in A$ for any $A \in \calA$. Suppose further, that for some family of constants $(c^\ixA)_{\ixA \in I} \subseteq \R_+ \cup \lbrace \infty \rbrace$ with
    \begin{align*}
        \pi^\ixA(t) < c^\ixA, \quad \forall t \in [0,T], \quad \forall \ixA \in I,
    \end{align*}
    the original market $((S^\ixA)_{\ixA\in I}, \frakF, \frakG, \sup_{\ixA \in I}c^\ixA)$ fulfils the \eqref{id:NoAsymptoticLpFreeLunch} condition. Then, the $\bfpi$-delayed market $((S^{\ixA, \bfpi})_{\ixA\in I}, \frakF, \frakG^\bfpi, T)$ is free of asymptotic $L^p$-free lunch.
\end{thm}

\begin{proof}
    Since the \eqref{id:NoAsymptoticLpFreeLunch} condition on $((S^\ixA)_{\ixA\in I}, \frakF, \frakG, \sup_{\ixA \in I}c^\ixA)$ is fulfilled, we know by the fundamental theorem of asset pricing (Theorem \ref{thm:FTAP}) that there exists a measure $\bbQ$ such that $\rmd\bbQ/\rmd\bbP \in L^q(\Omega, \calG, \bbP)$ and
    \begin{align}
        \forall \ixA \in A \text{ holds } \myCond{\bbQ}{S^\ixA_{u}}{\calF^A_t} = \myCond{\bbQ}{S^\ixA_{t}}{\calF^A_t} \text{ a.s.} \text{ with } A \in \calA,~ 0 \leq t \leq u \leq \textstyle\sup_{\ixA \in I}c^\ixA.\label{id:NonDelayedUnderlyinMartigale}
    \end{align}
    Fix any $A \in \calA$, $\ixA \in A$ and define $X = (X_t)_{t \in \R}$ by
    \begin{align*}
        X_t := \myCond{\bbQ}{S^\ixA_{t}}{\calF^A_t}, \quad \forall t \in [0,\textstyle\sup_{\ixA \in I}c^\ixA],
    \end{align*}
    which we conclude is an $\frakF^A$-martingale.
    
    For the sake of notation, we write $\pi \equiv \pi^\ixA$. Now, fix $0 \leq t \leq u \leq T$ and observe
    \begin{align*}
        \myCond{\bbQ}{S^\ixA_{\pi(u)}}{\calF^A_t} 
        &= \myCond{\bbQ}{
        \myCond{\bbQ}{
        S^\ixA_{\pi(u)}
        }{\calF^A_{\pi(u)}}
        }{\calF^A_t} 
        = \myCond{\bbQ}{
        X_{\pi(u)}
        }{\calF^A_t},
    \end{align*}
    due to the definition of $X$ and the fact that $\calF^A_{\pi(u)} \supseteq \calF^A_{t}$. Especially, we know $\calF^A_{\pi(t)} \supseteq \calF^A_{t}$ and therefore apply again the tower property of conditional expectation to obtain
    \begin{align*}
        \myCond{\bbQ}{
        X_{\pi(u)}
        }{\calF^A_t} = \myCond{\bbQ}{
        \myCond{\bbQ}{
        X_{\pi(u)}
        }{\calF^A_{\pi(t)}}
        }{\calF^A_t}.
    \end{align*}
    
    As a next step, we apply Doob's optional sampling theorem in the equation above to obtain $\E_{\bbQ}[{X_{\pi(u)}}|{\calF^A_{\pi(t)}}] = X_{\pi(t)}$. Since $\pi(t) \leq \pi(u)$ a.s. are $\frakI^\ixA$-stopping times and therefore $\frakF^A$-stopping times, we only have to verify that the martingale $X$ is càdlàg and closed for the application of the theorem. Right-continuity is indeed fulfilled due to the optional projection theorem, together with Assumption \ref{ass:Cadlag}. The optional projection theorem is applicable, since $\vert S^\ixA \vert$ is a $\bbQ$-sub-martingale and hence $L^1(\Omega, \calG, \bbQ)$-bounded by $\myE{\bbQ}{\vert S^\ixA_T \vert}< \infty$.
    
    Regarding closedness, the case $c^\ixA < \infty$ follows directly by redefining $X$ to its $\pi(u)$-stopped -- thus clearly closed -- version. Note that the stopped process is well-defined due to Assumption \ref{ass:Progressive}.
    
    Thus, assume $c^\ixA = \infty$, directly implying $\textstyle\sup_{\ixA \in I}c^\ixA = \infty$. Therefore, we know due to \eqref{id:NonDelayedUnderlyinMartigale} that $S^\ixA$ is closed under $\bbQ$ by $S^\ixA_\infty$. Moreover, due to Assumption \ref{ass:Lpintegrable} it follows that
    \begin{align*}
        \Vert S^\ixA_\infty \Vert_{L^1(\Omega, \calG, \bbQ)} \leq \Vert S^\ixA_\infty \Vert_{L^p(\Omega, \calG, \bbP)} \Vert \rmd\bbQ/\rmd\bbP \Vert_{L^q(\Omega, \calG, \bbP)} < \infty,
    \end{align*}
    where we used Hölder's inequality and the fact that $\Vert \rmd\bbQ/\rmd\bbP \Vert_{L^q(\Omega, \calG, \bbP)}$ is a finite constant by the definition of $\bbQ$.
    By the tower property we can rewrite $X_t := \myCond{\bbQ}{S^\ixA_{\infty}}{\calF^A_t}$ for all $t \in \R_+ \cup \lbrace \infty \rbrace$ and then Levy's 0-1-law directly implies that $X$ is closed.
    
    Combining the above results we receive 
    \begin{align*}
        \myCond{\bbQ}{S^\ixA_{\pi(u)}}{\calF^A_t} = \myCond{\bbQ}{
        X_{\pi(t)}}
        {\calF^A_t} = \myCond{\bbQ}{
        \myCond{\bbQ}{S^\ixA_{\pi(t)}}{\calF^A_{\pi(t)}}}
        {\calF^A_t} = \myCond{\bbQ}{
        S^\ixA_{\pi(t)}}
        {\calF^A_t},
    \end{align*}
    where we inserted the definition of $X$ and applied once more the tower property.
    
    We have shown that $\bbQ$ represents an equivalent martingale measure in the sense of Definition \ref{def:EMM} for the market $((S^{\ixA, \bfpi})_{\ixA\in I}, \frakF, \frakG^\bfpi)$. Note in particular, that the measure is well-defined in this setting, since we fixed $\calG \equiv \calG_\infty$ to especially achieve this stability under time shifts in $\frakG$. 
    
    Eventually, Theorem \ref{thm:FTAP} yields the desired result, being available on this market due to Lemma \ref{lem:PiDelayedTrading}.
\end{proof}

\begin{exa}[Insider trading revisited]
    The converse of Theorem \ref{thm:NAFLpWithExecDelay} does not hold in general. With the notation of the theorem, for any $h > 0$ it is possible to construct a market $((S^{\ixA, \bfpi})_{\ixA\in I}, \frakF, \frakG^\bfpi, 1)$ that is free of asymptotic $L^p$-free lunch, without the original market $((S^\ixA)_{\ixA\in I}, \frakF, \frakG, 1 + h)$ satisfying the \eqref{id:NoAsymptoticLpFreeLunch} condition. Subsequently, an exemplary market is demonstrated.
    
    One possible construction is analogue to the one in Example \ref{exa:InsiderTrading}. Let $B := (B_t)_{t \in [0,1 + 2h]}$ represent a $\bbP$-Brownian motion. We identify $S := (S_t)_{t \in [0,1 + h]}$ with $S_t = B_{t}$ for all ${t \in [0,1 + h]}$. Next, we model the flow of information. This is done by $\frakF := (\calF_t)_{t \in [0,1+h]} \equiv \frakG := (\calG_t)_{t \in [0,1+h]}$, where
    \begin{align*}
        \calF_t = \calG_t := \sigma(\lbrace B_u \rbrace_{u\in [0,t + h] }), \quad \forall t \in [0, 1 + h].
    \end{align*}
    It is clear that this market admits a free-lunch, by defining the strategy $m = 1$, $t_0 = 0$, $t_1 = h$, and $H_{t_1} := 2\myind{S_{h} \geq S_0} - 1$. Observe that $H_{t_1} \in \calF_0$, due to $\calF_0 \supseteq \sigma(\lbrace B_0, B_h \rbrace)$. The corresponding wealth process's terminal value is $\vert S_h - S_0 \vert$ and clearly constituting a free-lunch opportunity. 
    
    However, in the market $\pi$-delayed with $\pi(t) := t + h$ for all $t \in [0,1]$ asymptotic $L^p$-free lunch is absent. To be precise, we have $\frakG^\pi := (\sigma(\lbrace B_u \rbrace_{u\in [0,t + 2h] }))_{t \in [0,1]}$ and $S^\pi := (B_{t + h})_{t \in [0,1]}$ with $\calF_t = \sigma(\lbrace S^\pi_t\rbrace_{t \in [0,1]}, \lbrace B_t\rbrace_{t \in [0,h]})$. Since the information created by $B_t$ for $t \leq h$ is irrelevant in the market, we find a measure $\bbQ \in \calM^\infty(\bbP)$ implying the ($\mathrm{NAFL}_1$) condition on the original market. 
\end{exa}

\medskip
To conclude this section, we consider Theorem \ref{thm:NAFLpWithExecDelay} in the scenario where multiple brokers are available, each with an individual family of order execution delays. We commence by adding a strictly stronger delay to our setting. 
\begin{cor}[Superimposing $\pi$-delays]\label{cor:Superimposing1}
    Let $\bfpi = (\pi^\ixA)_{\ixA\in I}$, $\Tilde{\bfpi} = ({\Tilde{\pi}}^\ixA)_{\ixA\in I}$ be families of execution delays where for every $\ixA \in I$, $\frakI^\ixA := (\calI^\ixA_t)_{t \in \R_+}$ (respectively $\Tilde{\frakI}^\ixA := (\Tilde{\calI}^\ixA_t)_{t \in \R_+}$) is the affiliated delay information to $\pi^\ixA$ (respectively $\Tilde{\pi}^\ixA$). Assume that
    \begin{enumerate}[label = \roman*)]
        \item \label{cond:SI_A}$\pi^\ixA(t) \leq \Tilde{\pi}^\ixA(t)$ a.s. for all $\ixA \in I$, $t \in [0,T]$; 
        \item every element of $\bfpi$ is required to be progressive and continuous;
        \item $\Tilde{\bfpi}$ is bounded from above by a family of constants $(c^\ixA)_{\ixA \in I} \subseteq \R_+ \cup \lbrace \infty \rbrace$, i.e.
        \begin{align*}
            \Tilde{\pi}^\ixA(t) < c^\ixA, \quad \forall t \in [0,T], \quad \forall \ixA \in I;
        \end{align*}
        \item for all $\ixA \in I$ it holds $\frakI^\ixA \subseteq \Tilde{\frakI}^\ixA$, with $\Tilde{\frakI}^\ixA$ right-continuous; and
        \item\label{cond:Superimposing_DelayInfo} $(\Tilde{\calI}_{\pi^\ixA(t)})_{t \in \R_+} \subseteq \frakF^A$ whenever $\ixA \in A$ for any $A \in \calA$.
    \end{enumerate}
    Suppose further, that the ${\bfpi}$-delayed market $((S^{\ixA, \bfpi})_{\ixA\in I}, \frakF, \frakG^\bfpi, \sup_{\ixA \in I}c^\ixA)$ satisfies the \eqref{id:NoAsymptoticLpFreeLunch} condition. Then, the $\Tilde{\bfpi}$-delayed market $((S^{\ixA, \Tilde{\bfpi}})_{\ixA\in I}, \frakF, \frakG^{\Tilde{\bfpi}}, T)$ is free of asymptotic $L^p$-free lunch.
\end{cor}

\begin{proof}
    Lemma \ref{lem:PiDelayedTrading} ensures that $((S^{\ixA, \bfpi})_{\ixA\in I}, \frakF, \frakG^\bfpi, \sup_{\ixA \in I}c^\ixA)$ is a standard platonic market. On one hand, Lemma \ref{lem:InheritanceOfProgressiveness} indicates that the processes $(S^{\ixA, \bfpi})_{\ixA\in I}$ are progressive. On the other hand, Lemma \ref{lem:InheritanceOfCadlag} yields the right-continuity of these processes. Therefore,  Assumptions \ref{ass:Progressive} and \ref{ass:Cadlag} are fulfilled for the asset price processes $(S^{\ixA, \bfpi})_{\ixA\in I}$.
    
    With that, we can define the family of execution delays $\hat{\bfpi} = ({\hat{\pi}}^\ixA)_{\ixA\in I}$ via
    \begin{align*}
        \hat{\pi}^\ixA (t, \omega) := (\pi^\ixA)^{-1} \big( \Tilde{\pi}^\ixA(t, \omega), \omega \big), \quad t \in \R_+ \cup \lbrace \infty \rbrace,
    \end{align*}
    whereby we identify $(\pi^\ixA)^{-1}(t, \omega):= \inf \lbrace{s \in [0,\infty]}:{\pi^\ixA(s, \omega) \geq t}\rbrace$ with the convention that the infimum is equal to infinity if the set is empty. As in the proof of Theorem \ref{thm:RelationInformationExecutionDelay}, we may replace the inequality with an equality due to continuity of $\pi^\ixA$.
    
    We will now show that $\hat{\pi}^\ixA$ satisfies in fact the conditions of Definition \ref{def:ExecutionDelay} where  
    \begin{align*}
        \hat{\frakI}^\ixA := (\hat{\calI}^\ixA_t)_{t \in \R_+} \quad\text{with}\quad
        \hat{\calI}^\ixA_t := \Tilde{\calI}^\ixA_{\pi^\ixA(t)}, \quad  \forall  t \in \R_+
    \end{align*}
    is the corresponding delay information.
    \begin{enumerate}[label = \roman*), wide]
        \item Let $u \in \R_+$, then for fixed $t \in \R_+ \cup \lbrace \infty \rbrace$ we see that
        \begin{align*}
            \hat{\pi}^\ixA(t) \leq u \quad \Leftrightarrow \quad  {\pi}^\ixA (\hat{\pi}^\ixA(t)) \leq {\pi}^\ixA (u) \quad \Leftrightarrow \quad {\pi}^\ixA\big((\pi^\ixA)^{-1}( \Tilde{\pi}^\ixA (t))\big) \leq {\pi}^\ixA (u),
        \end{align*}
        due to monotonicity of $\pi^\ixA$ and infer $\pi^\ixA \circ (\pi^\ixA)^{-1}\equiv 1$ by additionally considering continuity of $\pi^\ixA$. Hence,
        \begin{align*}
            \lbrace \hat{\pi}^\ixA(t) \leq u \rbrace = \lbrace \Tilde{\pi}^\ixA(t) \leq {\pi}^\ixA (u) \rbrace \in \hat{\calI}^\ixA_u.
        \end{align*}
        The above holds true, since $\hat{\calI}_t := \Tilde{\calI}^\ixA_{\pi^\ixA(t)}$ and both, $\pi^\ixA(u)$ and $\Tilde{\pi}^\ixA(t)$ are $\Tilde{\frakI}$-stopping times. So, by basic properties of stopping times we infer 
        \begin{align*}
            \lbrace \Tilde{\pi}^\ixA(t) \leq {\pi}^\ixA (u) \rbrace \in \Tilde{\calI}^\ixA_{\pi^\ixA(u)}.
        \end{align*}
        \item Fix again some $t \in \R_+ \cup \lbrace \infty \rbrace$, then 
        \begin{align*}
            \hat{\pi}^\ixA (t) =  (\pi^\ixA)^{-1}(\Tilde{\pi}^\ixA (t)) \geq (\pi^\ixA)^{-1}({\pi}^\ixA ( t)) = t.
        \end{align*}
        Here, we used that continuity together with monotonicity of $\pi^\ixA$ yield that $\pi^\ixA \circ (\pi^\ixA)^{-1} \equiv 1$.
        \item Eventually, for $0 \leq t \leq u \leq \infty$ we see that 
        \begin{align*}
            \hat{\pi}^\ixA (t) =  (\pi^\ixA)^{-1}(\Tilde{\pi}^\ixA (t)) \leq (\pi^\ixA)^{-1}(\Tilde{\pi}^\ixA (u))  = \hat{\pi}^\ixA (u).
        \end{align*}
    \end{enumerate}
    We conclude that $\hat{\bfpi}$ is in fact a family of order execution delays that satisfy 
    \begin{align*}
        {S}^{\ixA, \bfpi}_{\hat{\pi}^\ixA(t)} = S^\ixA_{\pi^a \circ \hat{\pi}(t)}= S^{\ixA, \Tilde{\bfpi}}_{t}, \quad \forall t \in [0,T], \quad \forall \ixA \in I.
    \end{align*}
    
    Hence, all necessary conditions for Theorem \ref{thm:NAFLpWithExecDelay} are fulfilled, and so the desired result follows.
\end{proof}

Requirement \ref{cond:SI_A} of Corollary \ref{cor:Superimposing1} is somewhat unpleasant since it demands that the available brokers are strictly sorted with respect to their pace to execute market orders. Nonetheless, this provides a technique to verify the absence of $L^p$-free lunch across all brokers, even if they are not ordered in speed. The idea is to examine a hypothetical market determined by the minimum of all broker's order execution delays for $L^p$-free lunch. The following corollary summarizes the preceding. 

\begin{cor}\label{cor:Superimposing2}
    For $l \in \lbrace 1, ..., k \rbrace$, $k \in \N$, let $\bfpi_l = (\pi_l^\ixA)_{\ixA\in I}$ be families of order execution delays where for every $\ixA \in I$ the affiliated delay information to $\pi_l^\ixA$, $\frakI^{\ixA}_l \equiv \frakI^{\ixA} := (\calI^{\ixA}_t)_{t \in \R_+}$, coincides.\footnote{This is done for the sake of simplicity, a more extensive result requires precise handling of Condition \ref{cond:Superimposing_DelayInfo} in Corollary \ref{cor:Superimposing1}.} Further, let the family of constants $(c^\ixA)_{\ixA \in I} \subseteq \R_+ \cup \lbrace \infty \rbrace$ be such that
    \begin{align*}
        {\pi}_l^\ixA(t) < c^\ixA, \quad \forall t \in [0,T], \quad \forall \ixA \in I, \quad \forall l \in \lbrace 1, ..., k \rbrace.
    \end{align*}
    Moreover, assume that every element of $\bfpi_l$ is required to be progressive and right-continuous for all $l \in \lbrace 1, ..., k \rbrace$ and $({\calI}_{\pi_*^\ixA(t)})_{t \in \R_+} \subseteq \frakF^A$ whenever $\ixA \in A$ for any $A \in \calA$, where $\bfpi_* := (\pi_*^\ixA)_{\ixA \in I} := (\min_{l = 1}^k\pi_l^\ixA)_{\ixA \in I}$.

    Then, the ${\bfpi}_l$-delayed market $((S^{\ixA, \Tilde{\bfpi}_l})_{\ixA\in I}, \frakF, \frakG^{\Tilde{\bfpi}_l}, T)$ is free of asymptotic $L^p$-free lunch for all $l \in \lbrace 1, ..., k \rbrace$ if the ${\bfpi}$-delayed market $((S^{\ixA, \bfpi_*})_{\ixA\in I}, \frakF, \frakG^{\bfpi_*}, T = \sup_{\ixA \in I}c^\ixA)$ satisfies the \eqref{id:NoAsymptoticLpFreeLunch} condition.
\end{cor}

\begin{proof}
    The result follows directly from Corollary \ref{cor:Superimposing1}.
\end{proof}

\section*{Appendix}
\begin{restatable}[Inheritance of progressiveness under execution delays]{lem}{InheritanceOfProgressiveness}\label{lem:InheritanceOfProgressiveness}
    If Assumption \ref{ass:Progressive} is fulfilled and $\pi^\ixA$ is progressive for a family of execution delays $\bfpi := (\pi^\ixA)_{\ixA \in I}$ and every $\ixA \in I$, then $S^{\ixA, \bfpi}$ is progressively measurable with respect to $\frakG^\bfpi$ for any $\ixA \in I$.
\end{restatable}
\begin{proof}
Fix any $\ixA \in I$ and $t \in [0,T]$. We will to show that the mapping 
\begin{align*}
    \Tilde{S}: 
    \begin{array}{ccc}
        [0,t] \times \Omega & \longrightarrow & \R \\
        (u,\omega) & \longmapsto & S^{\ixA}_{\pi^\ixA(u, \omega)}(\omega)
    \end{array}
\end{align*}
is $\calB([0,t])\otimes \calG^\bfpi_t$ measurable. For this, consider some set $B \in \calB(\R)$. Observe that 
\begin{align*}
    &\mySet{(u, \omega) \in [0,t] \times \Omega}{S^{\ixA}_{\pi^\ixA(u, \omega)}(\omega) \in B} \\&\quad \quad \quad \quad  = \mySet{(u, \omega) \in [0,t] \times \Omega}{S^{\ixA}_{r}(\omega) \in B \text{ for some } r \in \R_+ \text{ with } \pi^\ixA(u, \omega) = r}
\end{align*}
and denote $(S^\ixA)^{-1}[B] =: B_1 \times G_1$. We know that $B_1 \in \calB(\R_+)$ and $G_1 \in \calG_t$ due to Assumption \ref{ass:Progressive}. Accordingly, $(\pi^\ixA)^{-1}[B_1] =: B_2 \times G_2 \in \calB([0,t]) \otimes \calG^\bfpi_t$ due to the progressiveness of $\pi^\ixA$. We claim that 
\begin{align*}
    \mySet{(u, \omega) \in [0,t] \times \Omega}{S^{\ixA}_{r}(\omega) \in B,~ r \in \R_+,~ \pi^\ixA(u, \omega) = r} = B_2 \times (G_1 \cap G_2).
\end{align*}
To show this, assume that $(u, \omega)$ satisfies $S^{\ixA}_{r}(\omega) \in B$ for some $r \in \R_+$ for that holds $\pi^\ixA(u, \omega) = r$. The former indicates that $(r, \omega) \in B_1 \times G_1$ and the latter implies $(u, \omega) \in (\pi^\ixA)^{-1}[\lbrace r \rbrace] \subseteq B_2 \times G_2$, due to $r \in B_1$. We conclude that $\omega \in G_1 \cap G_2$ and $u \in B_2$. 

For the converse, consider $(u, \omega) \in B_2 \times (G_1 \cap G_2)$. This directly implies that $r := \pi^\ixA(u, \omega) \in B_1$, by the definition of $B_2 \times G_2$. Hence, $(r, \omega) \in B_1 \times G_1$ and we conclude that $S^{\ixA}_{r}(\omega) \in B$. We have shown that the sets coincide. 

It is clear that $B_2 \times (G_1 \cap G_2) \in \calB([0,t])\otimes \calG^\bfpi_t$ due to $G_1 \in \calG_t \subseteq \calG_{\pi^\ixA(t)} \subseteq \calG^\bfpi_t$, thus we have shown the desired result.
\end{proof}

\begin{restatable}[Inheritance of right-continuity under execution delays]{lem}{InheritanceOfCadlag}\label{lem:InheritanceOfCadlag}
    If Assumption \ref{ass:Cadlag} is fulfilled and $\pi^\ixA$ is path-wise right-continuous for a family of execution delays $\bfpi = (\pi^\ixA)_{\ixA \in I}$ as well as every $\ixA \in I$, then $S^{\ixA, \bfpi}$ is càdlàg for any $\ixA \in I$.
\end{restatable}
\begin{proof}
For any fixed $\ixA \in I$ and $\omega \in \Omega$, the mappings $S^\ixA(\cdot)$ and $\pi^\ixA(\cdot)$ are right-continuous. Thus, the composition therefrom $S^\ixA \circ \pi^\ixA(\cdot)$ is right-continuous as well. 
\end{proof}

\bibliographystyle{unsrtnat}
\bibliography{references}

\begin{thebibliography}{10}
\providecommand{\natexlab}[1]{#1}
\providecommand{\url}[1]{\texttt{#1}}
\expandafter\ifx\csname urlstyle\endcsname\relax
  \providecommand{\doi}[1]{doi: #1}\else
  \providecommand{\doi}{doi: \begingroup \urlstyle{rm}\Url}\fi

\bibitem[Cuchiero et~al.(2020)Cuchiero, Klein, and
  Teichmann]{CuchieroKleinTeichmann20}
C.~Cuchiero, I.~Klein, and J.~Teichmann.
\newblock A fundamental theorem of asset pricing for continuous time large
  financial markets in a two filtration setting.
\newblock \emph{Theory Probab. Appl.}, 65\penalty0 (3):\penalty0 388--404,
  2020.
\newblock ISSN 1095-7219.
\newblock \doi{10.1137/S0040585X97T990022}.

\bibitem[Stricker(1990)]{Stricker90}
Christophe Stricker.
\newblock Arbitrage et lois de martingale.
\newblock \emph{Ann. Inst. H. Poincar\'{e} Probab. Statist.}, 26\penalty0
  (3):\penalty0 451--460, 1990.
\newblock ISSN 0246-0203.

\bibitem[Kabanov and Stricker(2006)]{KabanovStricker06}
Yuri Kabanov and Christophe Stricker.
\newblock The {D}alang-{M}orton-{W}illinger theorem under delayed and
  restricted information.
\newblock In \emph{In memoriam {P}aul-{A}ndr\'{e} {M}eyer: {S}\'{e}minaire de
  {P}robabilit\'{e}s {XXXIX}}, volume 1874 of \emph{Lecture Notes in Math.},
  pages 209--213. Springer, Berlin, 2006.
\newblock \doi{10.1007/978-3-540-35513-7_16}.

\bibitem[Cuchiero et~al.(2016)Cuchiero, Klein, and
  Teichmann]{CuchieroKleinTeichmann16}
C.~Cuchiero, I.~Klein, and J.~Teichmann.
\newblock A new perspective on the fundamental theorem of asset pricing for
  large financial markets.
\newblock \emph{Theory Probab. Appl.}, 60\penalty0 (4):\penalty0 561--579,
  2016.
\newblock ISSN 0040-585X.
\newblock \doi{10.1137/S0040585X97T987879}.

\bibitem[Delbaen and Schachermayer(1994)]{DelbaenSchachermayer94}
Freddy Delbaen and Walter Schachermayer.
\newblock A general version of the fundamental theorem of asset pricing.
\newblock \emph{Math. Ann.}, 300\penalty0 (3):\penalty0 463--520, 1994.
\newblock ISSN 0025-5831.
\newblock \doi{10.1007/BF01450498}.

\bibitem[Baldauf and Mollner(2020)]{BaldaufMollner20}
Markus Baldauf and Joshua Mollner.
\newblock High-frequency trading and market performance.
\newblock \emph{The Journal of Finance}, 75\penalty0 (3):\penalty0 1495--1526,
  2020.
\newblock \doi{10.1111/jofi.12882}.

\bibitem[Brolley and Cimon(2020)]{BrolleyCimon20}
Michael Brolley and David~A. Cimon.
\newblock Order-flow segmentation, liquidity, and price discovery: The role of
  latency delays.
\newblock \emph{Journal of Financial and Quantitative Analysis}, 55\penalty0
  (8):\penalty0 2555–2587, 2020.
\newblock \doi{10.1017/S002210901900067X}.

\bibitem[Hoffmann(2014)]{Hoffmann14}
Peter Hoffmann.
\newblock A dynamic limit order market with fast and slow traders.
\newblock \emph{Journal of Financial Economics}, 113\penalty0 (1):\penalty0
  156--169, 2014.
\newblock \doi{10.1016/j.jfineco.2014.04.002}.

\bibitem[Kyle and Lee(2017)]{KyleLee17}
Albert~S Kyle and Jeongmin Lee.
\newblock {Toward a fully continuous exchange}.
\newblock \emph{Oxford Review of Economic Policy}, 33\penalty0 (4):\penalty0
  650--675, 11 2017.
\newblock ISSN 0266-903X.
\newblock \doi{10.1093/oxrep/grx042}.

\bibitem[Pagnotta and Philippon(2018)]{PagnottaPhilippon18}
Emiliano~S. Pagnotta and Thomas Philippon.
\newblock Competing on speed.
\newblock \emph{Econometrica}, 86\penalty0 (3):\penalty0 1067--1115, 2018.
\newblock \doi{10.3982/ECTA10762}.

\end{thebibliography}

\end{document}